\documentclass[pra, showpacs, twocolumn, a4paper, nofootinbib]{revtex4}
\usepackage{amsmath}
\usepackage{verbatim}
\usepackage{bm}

\begin{document}

\newcommand{\ket}[1]{| #1 \rangle}
\newcommand{\bra}[1]{\langle #1 |}

\title{Multi-quantum eigenstates of a linear chain of coupled qubits}

\author{C.J. Mewton}
\author{Z. Ficek}
\affiliation{Department of Physics, School of Physical Sciences,
The University of Queensland, Brisbane, QLD, Australia 4072}
\date{\today}

\begin{abstract}
We present a technique to identify exact analytic expressions for the 
multi-quantum eigenstates of a linear chain of coupled qubits. 
A choice of Hilbert subspaces is described which allows an exact solution 
of the stationary Schr\"{o}dinger equation without imposing periodic boundary 
conditions and without neglecting end effects, fully including the 
dipole-dipole nearest-neighbor interaction between the atoms. The treatment 
is valid for an arbitrary coherent excitation in the atomic system, any number 
of atoms, any size of the chain relative to the resonant wavelength and 
arbitrary initial conditions of the atomic system. The procedure we develop 
is general enough to be adopted for the study of excitation in 
an arbitrary array of atoms including spin chains and one-dimensional 
Bose-Einstein condensates.
\end{abstract}


\maketitle

It is well established that a large scale quantum computation will 
require a large number of strongly coupled atoms (qubits) and in 
particular multi-photon excitations to create large quantum 
superpositions of different quantum states of the system 
\cite{bib:Nielsen2000}.
Naturally, this means that an operation of such systems will require 
collective excitations involving more than one photon. 
Although collective excitations in multi-atom systems have been 
studied in the past, it was typically done in the context of radiation 
properties of a large number of atoms randomly distributed in space, 
such as in a gas cell~\cite{bib:Ficek2002}. 
There have been many exact studies of collective eigenstates of spatially 
ordered systems, such as Heisenberg chains~\cite{bib:Bethe1931,bib:Lieb1961,
bib:Brennen2004}. 
However, these studies suffer from one common drawback: they are limited to 
finding the ground state or one-photon excited states in the limit of a large 
number of atoms, or are subject to periodic boundary conditions. 

With the recent progress in trapping and cooling of a small number of atoms
or ions, the attention has been drawn towards systems that are comprised 
of a small number of atoms having a definite geometrical arrangement, 
as dictated by the confining field of a hypothetical atom trap 
\cite{bib:Leibfried2003}. A number of analytical studies have been 
performed on systems composed of $N$ atoms confined to fixed 
positions, with a particular interest in ring
arrangements~\cite{bib:Freedhoff2004}, perhaps as
they have periodic boundary conditions which makes the eigenstates
easier to obtain.  However, exact analytic studies, which give 
explicit forms of the eigenstates, have been performed on 
systems containing only two and three atoms \cite{bib:Freedhoff2004,%
 bib:Richter1982,bib:Freedhoff1986}.
More general results valid for arbitrary $N$ have also been 
given~\cite{bib:Foldi2002,bib:Hammer2004}, but are limited
to one-photon excitations only. More exotic shapes, such as diamond
structures \cite{bib:Rudolph2004}, have also been
investigated, but the calculations were limited to two-photon 
excitations and small numbers of atoms. While these schemes successfully 
demonstrate the situation for obtaining one- or two-photon-excited states 
in $N$-atom systems, the analysis for the excitation of such a system
by an arbitrary number of photons, giving a multi-quantum eigenstate,
is unknown.

Other treatments of these systems stem from statistical physics, where
properties are extracted from the partition function, which is determined
by the eigenvalues of the system.
The partition function can provide a witness to
entanglement \cite{bib:Wiesniak2005}, it does not, however, give information
about the energy eigenstates of the system.
  These states may have applications in quantum information and need
 to be quantified in order to perform quantum
computation.
  This forms the motivation of this work to derive the 
explicit analytic expressions for the multi-quantum eigenstates
 of a linear chain of $N$ atoms that can be applied to 
a variety of settings. 
The availability of these eigenstates is highly 
advantageous because it provides a convenient ground for rigorous
examination of entangled properties of a particular arrangement of 
atoms and facilitates the explicit study of the stability of 
entangled states, something which can be obscured if we remove this 
generality.  It may also provide an interesting tool to study entanglement
creation in 1D Bose-Einstein condensates \cite{bib:Sorensen2001}.
We cover the case of the linear 
chain as although it is one of the simplest structures that can be 
constructed in an atom trap, it has not been solved exactly while 
imposing the generality that we demand here. There is also a considerable 
interest in analyzing multi-atom systems to realize a phase gate operation
\cite{bib:Beige2000}, in which we do not impose periodic boundary
conditions and include the end effects.

In this paper, we study the creation of multi-quantum eigenstates 
in a linear chain of $N$ identical, equally spaced and confined to fixed
positions, two-level atoms each interacting only with its nearest neighbors 
through the dipole-dipole interaction. This model might be realized in 
practice by placing the atoms in a tightly confining linear trap, 
an optical lattice or in an atomic chip~\cite{bib:Jessen1996, bib:Folman2000}. 
The Hamiltonian for the linear chain of atoms is given~by
\begin{equation}
\hat{H} = \hat{H}_0 + \hat{V} = \hbar \omega_0 \sum_{i=1}^{N} \hat{S}_i^z
+\hbar \Omega \sum_{ \substack{i,j=1\\|i-j|=1} }^N
\hat{S}_i^+ \hat{S}_j^- ,
\label{eq:hamiltonian}
\end{equation}
where $\hat{H}_0$ is the interaction-free Hamiltonian and 
$\hat{V}$ is the dipole-dipole interaction between the atoms.

In Eq.~(\ref{eq:hamiltonian}), $\omega_0$ is the transition frequency 
of a two-level atom in isolation, $\hat{S}_i^z$ is the energy operator of 
the $i$-th atom, and $\hat{S}_i^+$, $\hat{S}_i^-$ are the raising and
lowering operators for the $i$-th atom, respectively.  
The dipole-dipole interaction parameter $\Omega \equiv \Omega(r_{ij})$
depends on the distance $r_{ij}$ between adjacent atoms in the linear 
chain and the polarization of the atomic dipole moments relatively 
to the interatomic axis~\cite{bib:Ficek2002}.

To find the multi-quantum energy states of the system, 
we have to solve the stationary Schr\"{o}dinger equation with the 
Hamiltonian $\hat{H}$. We propose a procedure which, despite of the 
complexity of the problem, facilitates an exact analytic solution 
of the Schr\"{o}dinger equation valid for an arbitrary number of 
excitations~$M$. The interaction-free Hamiltonian $\hat{H}_0$ has 
$N+1$ energy levels of energies $E_0^{(M)} = M \hbar \omega_0$, 
where $M=0,1,2,\ldots,N$. Since there are $N$ two-level atoms, 
it follows that $\hat{H}_0$ operates on a  
$2^N$-dimensional Hilbert space~$W$.  We can index the atoms in the
chain by the numbers $1$ to $N$; collective excitations the system can then
be represented using the ket $\ket{k_1, \ldots, k_M}$, where the non-zero
integers $k_1, \ldots, k_M$ denote the indices of the atoms which are in
their excited state:
\begin{equation}
\ket{k_1, \ldots, k_M} = \hat{S}_{k_1}^+ \cdots \hat{S}_{k_M}^+ \ket{0} ,
\end{equation}
where $\ket{0}$ denotes the ground state of the collective system, and is
nondegenerate.  The ground states $\ket{0}$ and excited states
$\ket{k_1, \ldots, k_M}$ form a basis which spans $W$.
The maximally excited interaction-free energy eigenstate, i.e.
the collective state where all $N$ atoms are in their upper state,
is then represented by $\ket{1,2, \ldots, N}$ in our notation.

Under the action of $\hat{H}_0$, a particular energy level $M$ has a 
degeneracy factor of $N!/(N-M)!M!$, which is equal to the number 
of ways $N$
atoms can be divided into two groups, with $M$ atoms excited and
$N-M$ atoms in their ground states.  However, while $\hat{H}$ acts on
the same Hilbert space~$W$, it leads to a significant splitting of the 
degeneracies of the collective system.

We observe that the process of finding the eigenstates of $\hat{H}$ may be
simplified by noting that $\hat{H}_0$ and $\hat{V}$ commute. Since
\begin{eqnarray}
 \frac{[\hat{H_0}, \hat{V}]}{\omega_0 \Omega \hbar^2 } = 
 \sum_{ \substack{i,j, \, k=1\\|j-k|=1} }^N \left(
 S_j^+ [ S_i^z, S_k^- ] + [ S_i^z, S_j^+ ]  S_k^- \right) ,
\end{eqnarray}
and using the relations $[S_i^z, S_j^\pm] = \pm S_i^\pm
\delta_{ij}$, we can easily see that $[ \hat{H_0}, \hat{V} ] = 0$.
This result, coupled with the fact that $\hat{H_0}$ and $\hat{V}$
are Hermitian means that the operators $\hat{H_0}$ and $\hat{V}$
share a complete set of orthonormal eigenstates $\{ \ket{\phi_i}
\}$.  Since $\hat{H}$ is just the sum of these two operators, and
$\{ \ket{\phi_i} \}$ is a complete set, it follows that $\hat{H}$,
$\hat{H}_0$ and $\hat{V}$ share the same complete set of
orthonormal eigenstates $\{ \ket{\phi_i} \}$.  The $\hat{H_0}$
operator allows us to partition the state vector space $W$ into
$N+1$ subspaces $W^j$:
\begin{equation}
W = \bigoplus_{j=0}^{N} W^j,
\end{equation}
where the subspace $W^j$ corresponds to all states which give an
eigenvalue of $j \hbar \omega$ when operated on by $\hat{H_0}$,
and is of dimensionality $N!/(N-j)!j!$.  In this way, we see that
$\hat{H}_0$ is an automorphism $T(\hat{H}_0) : W^j \rightarrow
W^j$ for all $W^j$.  Since $\hat{H}$ and $\hat{V}$ commute with
$\hat{H}_0$, they must be mappings of the form $W^j \rightarrow
W^j$ for all subspaces $W^j$.  Thus, to find the energy
eigenstates of the system, we can solve the 
Schr\"{o}dinger equation for $\hat{H}$ for each subspace $W^j$
instead of $W$.  However, all states in a particular $W^j$ are
eigenstates of $\hat{H}_0$, so we can neglect this operator and
simply solve the Schr\"{o}dinger equation for $\hat{V}$.  
To accomplish this, we take the most general state vector one can 
form in the subspace $W^M$:
\begin{eqnarray}
\ket{\psi^{(M)}} & = & \sum_{k_1 < \cdots < k_M}
 C(k_1, \ldots, k_M) \ket{k_1, \ldots, k_M},
\label{eq:psiM}
\end{eqnarray}
where the summation runs through all $M$-tuples $(k_1, \ldots, k_M)$ such
that $1 \leq k_1 < \cdots < k_M \leq M$.
To then obtain the energy eigenstates, we solve
\begin{equation}
\hat{V} \ket{\psi^{(M)}} = \Delta E \ket{\psi^{(M)}} ,
\label{eq:dipoleEigenvalue}
\end{equation}
where $\ket{\psi^{(M)}} \in W^M$ and $\Delta E$ is the energy
level shift associated with the dipole-dipole interaction
$\hat{V}$.  The eigenvalue associated with $\hat{H}$ is then given
by
\begin{eqnarray}
\hat{H} \ket{\psi^{(M)}} & = & \hat{H}_0 \ket{\psi^{(M)}}
 + \hat{V} \ket{\psi^{(M)}} \nonumber \\
& =& (E^{(M)}_0 + \Delta E) \ket{\psi^{(M)}}.
\end{eqnarray}

%
%

We assume that the energy eigenstates of the interaction
Hamiltonian are of the form (\ref{eq:psiM}) and constitute a complete set.  
In order to determine the
coefficients $C(k_1, \ldots, k_M)$, we substitute (\ref{eq:psiM})
into Eq. (\ref{eq:dipoleEigenvalue}), and find
the following multi-term recurrence relation relating the
coefficients $C(k_1, \ldots, k_M)$ to the energy eigenvalue 
shift~$\Delta E$:
\begin{eqnarray}
\Delta E C(k_1, \ldots, k_M) & = & \hbar \Omega \sum_{j=1}^{M}
 [ C(k_1, \ldots, k_j +1, \ldots, k_M) \nonumber \\
& &  + C(k_1, \ldots, k_j -1, \ldots, k_M) ],
 \label{eq:recurrenceRelation}
\end{eqnarray}
where it is assumed that the summation in the above expression must not 
include terms which have an invalid value for $k_j$, i.e. are not generated by
Eq. (\ref{eq:dipoleEigenvalue}).  There are only three such cases:
(a) when $k_1$ is equal to zero or (b)
$k_M = N+1$ and thus the index cannot refer to an actual atom, and
(c) when two of the $k_j$ are equal, which would rule out the
state from being an eigenvector of $\hat{V}$.

Equation (\ref{eq:recurrenceRelation}) without such a restriction on summation
can be solved easily, so we prescribe the following method: we allow
invalid terms to be included in Eq. (\ref{eq:recurrenceRelation}), on the
condition that they must vanish.  This means that we now regard cases
(a), (b), and (c) as conditions under which a coefficient must vanish.

By inspection, we see that Eq. (\ref{eq:recurrenceRelation}) 
resembles a pattern that is seen in sine or cosine functions $f(x)$,
for $f(x+1) + f(x-1) \propto f(x)$.  As a trial solution, we thus
consider the general form
\begin{eqnarray}
C(k_1, \ldots, k_M) & = & \sum_{a, \cdots, d} K_{a \cdots d}
 f_a( \alpha_a k_1 + A_a) \cdots \nonumber \\
& & \quad \cdots f_d( \delta_d k_M + D_d),
\end{eqnarray}
where $\alpha_a, \ldots, \delta_d$ and $A_a, \ldots, D_d$ are
arbitrary constants along with $K_{a \cdots d}$, and the summation range
for the $M$ indices $a, \ldots, d$ ranges from $1$ to $M$.
By operating
on the $k_j$ individually by sine or cosine functions, this expansion can be
made fully consistent with Eq. (\ref{eq:recurrenceRelation}) using further
restrictions which now follow.

The only way that condition (c) can be satisfied is if $K_{a
\cdots d}$ is antisymmetric in any two pairs of indices.
Condition (a) then implies that either all the $f_i$ are sine
functions with the offsets $A_i, \ldots, D_d$ equal to zero or the
$f_i$ are cosine functions where $A_i, \ldots, D_d$ are all equal
to $\pi / 2$.  These solutions are not independent -- they are
equal up to a multiplicative constant, so we choose the sine
function.  The antisymmetry property also requires that $\alpha_i
= \cdots = \delta_i$. We then have
\begin{equation}
C(k_1, \ldots, k_M) = K \epsilon_{a \cdots d}
 \sin ( \alpha_a k_1 ) \cdots \sin ( \alpha_d k_M ),
\end{equation}
where $\epsilon_{a \cdots d}$ is the permutation (Levi-Civita)
symbol and summation is implied over the repeated $M$ symbols $a,
\ldots, d$ from $1$ to $M$.  The constant $K$ will be set equal 
to unity from now on as it is not constrained by any conditions.

For condition (b) to be satisfied, and owing to the permutations
of the $\alpha_i$, the constants $\alpha_1, \ldots, \alpha_M$ must
equal $g_1 \theta, \ldots, g_M \theta$, where $\theta = \pi /
(N+1)$ and the $g_i$ are integers.  Our solution now becomes
\begin{equation}
C(k_1, \ldots, k_M) = \epsilon_{a \cdots d}
 \sin ( g_a k_1 \theta ) \cdots \sin ( g_d k_M \theta ),
\end{equation}
or equivalently
\begin{equation}
C(k_1, \ldots, k_M) = \epsilon_{a \cdots d}
 \sin ( g_1 k_a \theta ) \cdots \sin ( g_M k_d \theta ).
\label{eq:coefficient}
\end{equation}

We have thus derived the coefficients apart from the arbitrary
integers $g_1, \ldots, g_M$.  Clearly, no two $g_i$ can be equal,
for the $C(k_1, \ldots, k_M)$ would vanish due to the presence of
the permutation symbol.  Also, to avoid duplicate cases, we
require for non-trivial eigenstates that
\begin{equation}
g_1 < \cdots < g_M.
\end{equation}
If a particular $g_i$ is equal to $N+1$, the coefficient $C(k_1,
\ldots, k_M)$ vanishes; therefore all the $g_i$ must be less than
or greater than $N+1$.  The difference between
these two cases amounts to multiplying Eq. (\ref{eq:coefficient})
by a minus sign on the right-hand side.  Since
such a factor is irrelevant, we impose the restriction that $g_i
\le N$ for all $g_i$.  Due to the properties of the sine function,
this implies also that $g_i \ge 1$ for all $g_i$.  Thus we can
state constraints on the $g_i$:
\begin{equation}
1 \le g_1 < \cdots < g_M \le N .\label{eq:g-constraints}
\end{equation}
Now that we have determined the coefficients of the eigenstates
(\ref{eq:psiM}), we are in a position to
determine the energy eigenvalue shift.  
Using (\ref{eq:recurrenceRelation}), we have
\begin{eqnarray}
&& \Delta E C(k_1, \ldots, k_M) \nonumber \\
&& = \hbar \Omega \sum_{j=1}^{M}
 2 \cos (g_c \theta )\left[\epsilon_{a \cdots c \cdots e}
\sin  (g_a k_1 \theta )\cdots\right.  \nonumber \\
& & \left. \cdots \sin (g_c k_j \theta )\cdots \sin (g_e k_M \theta )\right] ,
\end{eqnarray}
where we have converted pairs of terms of the form
$\sin g_c ( k_j + 1) \theta $ and $\sin g_c ( k_j - 1) \theta$
into the product $2 \cos g_c \theta \sin g_c k_j \theta$.  We
repeat this for each sine pair, giving the end result
\begin{eqnarray}
\Delta E C(k_1, \ldots, k_M) & = &
2 \hbar \Omega C(k_1, \ldots, k_M) \nonumber \\
& & \times \sum_{i=1}^M \cos (g_i \theta ) .
\end{eqnarray}

The energy eigenvalue shift is thus given by
\begin{equation}
\Delta E = 2 \hbar \Omega \sum_{i=1}^M \cos (g_i \theta ) .
\end{equation}

%
%

We now collect the results to give the energy eigenstates and
eigenvalues for the system (\ref{eq:hamiltonian}), valid for an
arbitrary number of atoms $N$ and an arbitrary number of
excitations $M$.  The unnormalized energy eigenstates 
of (\ref{eq:hamiltonian}) are given~by
\begin{eqnarray}
&& \ket{\psi^{(M)}_{g_1 \cdots g_M}} = \sum_{k_1 < \cdots < k_M}
 C^{k_1 \cdots k_M}_{g_1 \cdots g_M} \ket{k_1, \ldots, k_M},
 \label{eq:energyEigenstates}
\end{eqnarray}
with corresponding eigenvalues
\begin{eqnarray}
E^{(M)}_{g_1 \cdots g_M} & = & M \hbar \omega_0
 + 2 \hbar \Omega \sum_{i=1}^M \cos (g_i \theta )
\end{eqnarray}
where
\begin{equation}
C^{k_1 \cdots k_M}_{g_1 \cdots g_M} = \epsilon_{a \cdots d}
\sin ( g_a k_1 \theta ) \cdots \sin ( g_d k_M \theta ) ,
\end{equation}
and $\theta = \pi / (N+1)$. The parameter
$M$ equals the number of excited atoms in a particular eigenstate,
and thus forms one of the quantum numbers that enumerate the
energy eigenstates.  The other quantum numbers are the $M$ numbers
$g_i$, which are subject to the constraint~(\ref{eq:g-constraints}).
There are $N!/(N-M)!$ ways of choosing values for the~$g_i$ so
that for all $g_i$, $1 \le g_i \le N$.  However, the
constraint~(\ref{eq:g-constraints}) forces us to divide the number
of combinations by the
number of ways one can re-label the $g_i$, which is~$M!$. Hence
there are $N!/(N-M)!M!$ energy eigenstates associated with a given 
value of $M$.  Note
that the number of the states for a given $M$ is the 
same as it was in the interaction-free system mentioned earlier.

To fit the ground eigenstate ($M=0$) into the 
solution~(\ref{eq:energyEigenstates}), we observe that $C^{k_1 \cdots
k_M}_{g_1 \cdots g_M}$ has no~$g_i$'s and no~$k_j$'s for the case
$M=0$, so we define $C^{k_1 \cdots k_M}_{g_1 \cdots g_M} = C$, an
arbitrary constant which may be set equal to unity.  Clearly
the summation sign in (\ref{eq:energyEigenstates}) disappears in
this case, and the only ket present is defined to be that corresponding to the
ground state of the system, $\ket{0}$.

One can see the step which enabled us to determine the exact 
analytic expressions for the multi-quantum eigenstates. 
The recurrence relation derived from the
eigenvalue equation had restrictions on its summation range; however, it
is clear that there is a simple pattern in the summation formula.  We
completed the pattern by including invalid terms, and created boundary
conditions which made these extra terms vanish.  Other treatments handle
the problem of the restrictions on the original summation by imposing
periodic boundary conditions.  However, the latter treatment implicitly
converts the linear chain into a ring -- a different physical system.  That
the derived states are eigenstates of a linear chain Hamiltonian is the
justification of our approach.

As an illustration of our general solution, we discuss two
specific examples, $M=1$ and $M=2$, corresponding to single- and
two-photon excitation of the $N$-atom system.  The general
eigenstates (\ref{eq:energyEigenstates})  for $M=1$ simplify~to
\begin{equation}
\ket{\psi^{(1)}_{g_{1}}} = \sum_{k=1}^{N} \sin ( g_1 k \theta ) \ket{k} ,
\end{equation}
where $1 \le g_1 \le N$ is a single quantum number differentiating
between $N$ states.  The associated eigenvalues~are
\begin{equation}
E^{(1)}_{g_1} = \hbar \omega_0 + 2 \hbar \Omega \cos g_1 \theta.
\end{equation}
For these states, we can see an interesting property in the small
sample model, where the interatomic separation is much smaller
than transition wavelength \cite{bib:Dicke1954}.  For the states
where $g_1$ is even, the
transition dipole moment $\bra{\psi^{(1)}_g} \hat{\bm{\mu}}
\ket{0}$, where $\ket{0}$ is the ground state, vanishes.  The
reason for this is that the coefficient of $\ket{k}$ is the
negative of the coefficient of $\ket{N+1-k}$ in the above
eigenstate expansion for such states.  The sum of these
coefficients then gives zero for the transition dipole moment.
Thus $g_1$-even states do not spontaneously decay to the ground
state.

For $M=2$, the eigenstates are of the form
\begin{eqnarray}
\ket{\psi^{(2)}_{g_1 g_2}} & = & \sum_{k_1=1}^{N-1} \sum_{k_2 =
k_1 + 1}^{N}
 [ \sin ( g_1 k_1 \theta ) \sin ( g_2 k_2 \theta ) \nonumber \\
& & - \sin ( g_1 k_2 \theta ) \sin ( g_2 k_1 \theta ) ] \ket{k_1,
k_2} ,
\end{eqnarray}
with energies
\begin{equation}
E^{(2)}_{g_1 g_2} = 2 \hbar \omega_0
 + 2 \hbar \Omega ( \cos g_1 \theta + \cos g_2 \theta ) .
\end{equation}
For this case, $1 \le g_1 < g_2 \le N$, giving a total of $N(N-
1)/2$ eigenstates.  This is the first excitation level where the
antisymmetric property of the $C^{k_1 \cdots k_M}_{g_1 \cdots
g_M}$ appears. For a small number of atoms, $N=3$, the above 
eigenstate and eigenvalue examples
agree (apart from the normalization constant) with the
case of the isosceles triangle~\cite{bib:Freedhoff2004}
flattened into a straight line, with the dipole-dipole
interaction between the two atoms at the ends of the line
removed.

%
%

In summary, we have proposed a simple and robust method to derive exact 
expressions for the multi-quantum states of 
a linear chain of $N$ trapped atoms interacting with its nearest 
neighbors through the dipole-dipole interaction. The method is based on 
a technique of converting the eigenvalue equation into a specific recurrence 
relation, rather than the usual technique of diagonalizing a matrix, the 
latter being more suited to the case where the number of atoms is not
arbitrary.

\end{document}